\documentclass{article}

\usepackage{microtype}
\usepackage{graphicx}
\usepackage{subfigure}
\usepackage{booktabs} %
\usepackage{comment}

\usepackage{hyperref}

\usepackage[accepted]{synsml2023}

\usepackage{amsmath}
\usepackage{amssymb}
\usepackage{mathtools}
\usepackage{amsthm}
\usepackage{xspace}

\usepackage[capitalize,noabbrev]{cleveref}

\theoremstyle{plain}

\theoremstyle{definition}

\theoremstyle{remark}

\newcommand{\ainuc}{NuCLR\xspace} %

\usepackage[textsize=tiny]{todonotes}

\begin{document}

\twocolumn[

\synsmltitle{NuCLR: Nuclear Co-Learned Representations}

\synsmlsetsymbol{equal}{*}

\begin{synsmlauthorlist}
\synsmlauthor{Ouail Kitouni*}{iaifi}
\synsmlauthor{Niklas Nolte*}{iaifi}
\synsmlauthor{Sokratis Trifinopoulos}{iaifi}
\synsmlauthor{Subhash Kantamneni}{iaifi}
\synsmlauthor{Mike Williams}{iaifi}
\end{synsmlauthorlist}
\synsmlcorrespondingauthor{Mike Williams}{mwill@mit.edu}

\synsmlaffiliation{iaifi}{Institute for Artificial Intelligence and Fundamental Interactions (IAIFI), MIT, USA}

\synsmlkeywords{Machine Learning}

\vskip 0.3in
]

\printAffiliationsAndNotice{\synsmlEqualContribution} %

\begin{abstract}
We introduce Nuclear Co-Learned Representations (\ainuc), a deep learning model that predicts various nuclear observables, including binding and decay energies, and nuclear charge radii. 
The model is trained using a multi-task approach with shared representations and obtains state-of-the-art performance, 
achieving levels of precision that are crucial for understanding fundamental phenomena in nuclear (astro)physics. %
We also report an intriguing finding that the learned representations of \ainuc exhibit the prominent emergence of 
crucial aspects of the nuclear shell model, namely the shell structure, including the well-known magic numbers, and the Pauli Exclusion Principle.
This suggests that the model is capable of capturing the underlying physical principles, and that our approach has the potential to offer valuable insights into nuclear theory.

\end{abstract}

\section{Introduction}

The nucleus is the incredibly small and dense region at the center of an atom, consisting of protons and neutrons bound together by the strong nuclear force.
Despite having been discovered over a century ago---and the fundamental theory of the strong nuclear force (quantum chromodynamics, QCD) having been discovered 50 years ago---we still lack a precise quantitative understanding of nuclear physics.
In principle all nuclear properties are calculable from the theory of QCD, though in practice such calculations are intractable for all but the smallest few nuclei.

Qualitatively, a major breakthrough in understanding nuclei came just after World War II with the development of the {\em nuclear shell model} by Goeppert-Mayer and Jensen  (Nobel Prize in Physics, 1963), which analogous to the atomic shell model describes the structure of the nucleus in terms of quantum energy states.
A key component of this model is the Pauli Exclusion Principle, that no two identical fermions, {\em e.g.}\ nucleons (protons or neutrons), can occupy the same quantum state.
Since nucleons have spin 1/2, two such particles can occupy the same energy state, one with spin up and the other with spin down, leading to even numbers of protons and neutrons being preferred (more strongly bound).
As in the atomic shell model, the most stable states are those with filled shells.
Therefore, when adding nucleons to a nucleus, there are {\em magic numbers} where the binding energy of the next nucleon is substantially less than the previous one, {\em i.e.}\ there are configurations that are much more tightly bound due to having a filled shell.
The shells for protons and neutrons are filled independently since they are distinguishable particles; however, the number of protons does affect the neutron quantum states and {\em vice versa}.
There are well known magic numbers such as 2, 8, 20, 28, 50, and 82 for both proton and neutron number, $Z$ and $N$, respectively, and in addition $N=126$. %

\begin{figure*}[h!]
    \centering
\includegraphics[width=0.49\textwidth]{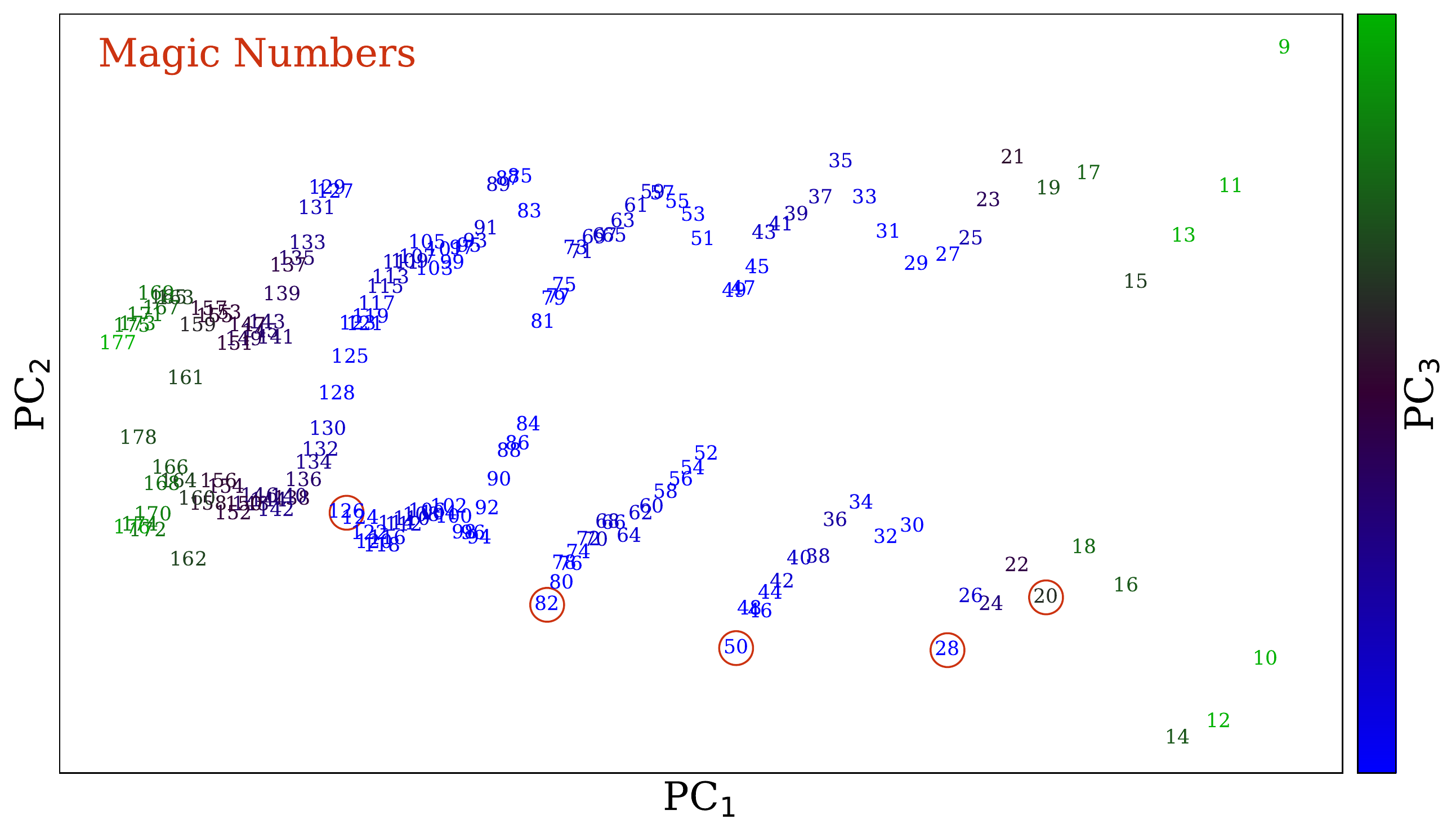}  
\includegraphics[width=0.49\textwidth]{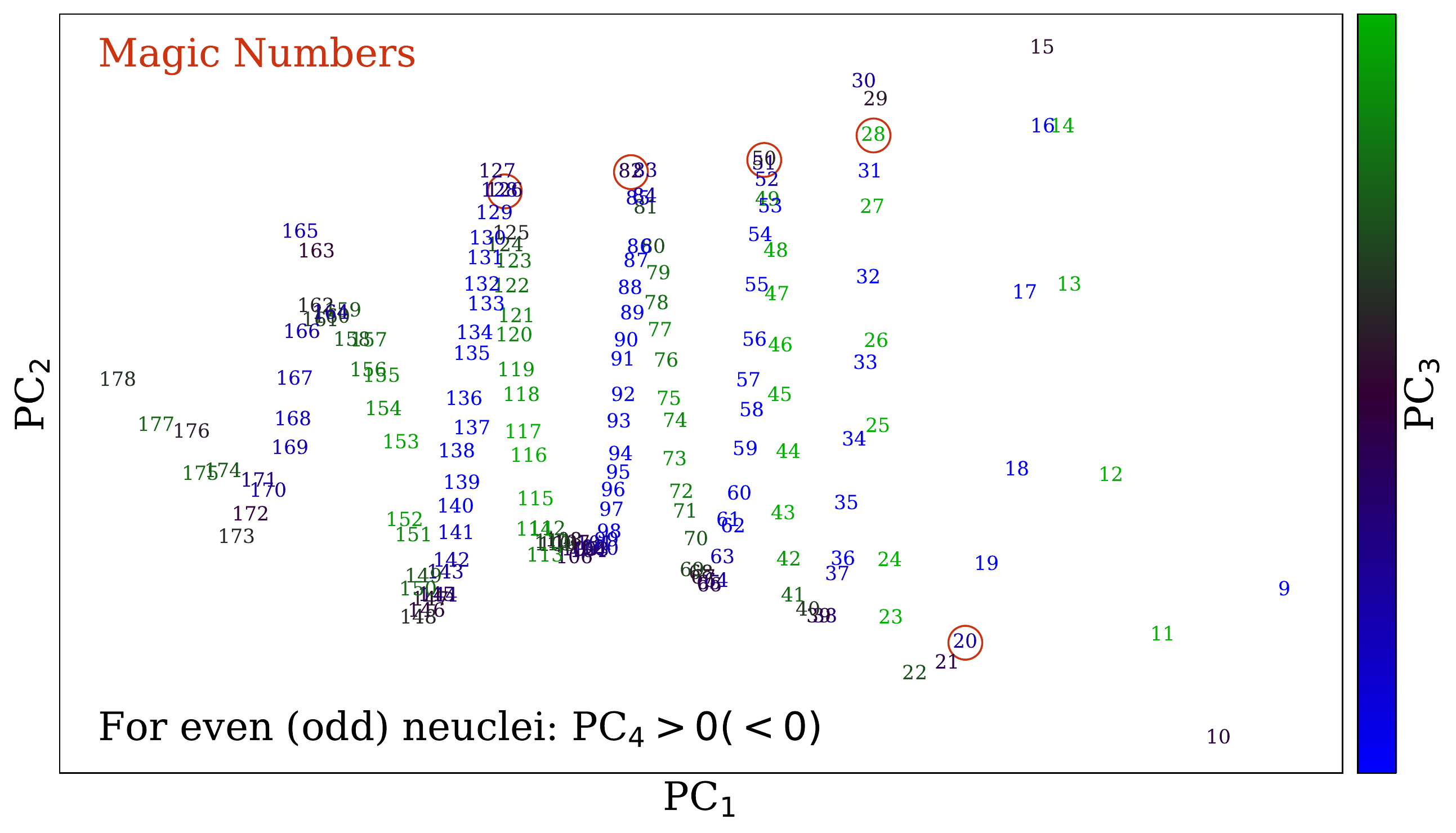}   \vspace{-0.135em}
    \caption{Most important principal components of the neutron embedding representations from (left) early and (right) late in the training. 
    The left panel shows that the most crucial aspects of the nuclear shell model, 
    namely the shell structure, including the magic numbers, and the Pauli Exclusion Principle, 
    arise already early in the training. (The Pauli principle is evident with the even (odd) numbers represented as negative (positve) values of PC2. The ends of the even chains of numbers are the magic numbers where each nuclear shell become full.)
    In the right panel, the even-odd split now occurs in PC dimension 4, hence is not shown. The shell structure has grown into 3-dimensional spirals, with the largest 4 magic numbers all occurring at local maxima in PC2 and each shell represented as one revolution around an approximately conic surface. Interpreting this spiral structure is ongoing work. 
    }
    \label{fig:nuc-emb}
\end{figure*}

A major goal of the modern nuclear physics community is to accurately predict nuclear properties, among which binding energies %
and charge radii are some of the most important.
Traditionally, analytical models have been used, motivated by the known physics---but simplified to make the calculations tractable---and augmented with empirical input to improve their agreement with data~\cite{Goriely:2001zz,Bender:2003jk,Geng:2005yu,Moller:2012pxr,Wang:2014qqa}.\footnote{With an increase in computational power nuclear {\em ab initio} methods have recently gained momentum~\cite{Navratil:2009ut,Hagen:2013nca,Ekstrom:2015rta,Novario:2021low}.}
The most accurate of these models, referred to as WS4~\cite{Wang:2014qqa}, is able to predict nuclear binding energies and charge radii with a precision of about 300\,keV and 0.02\,fm, respectively.  
More recently, machine learning (ML) models have received some attention due to their superior performance potential.
Thus far, most ML models were designed for a \emph{single-task learning} (STL) purpose, namely predicting binding energies~\cite{Niu:2018csp,Wu:2021hil,Niu:2022gwo,Wu:2022nnc} 
or charge radii~\cite{Utama:2016tcl,Wu:2020bao,Dong:2021aqg,Ma:2022yrj}.
An exception is the kernel ridge regression model of \cite{Wu:2022nnc} which utilizes \emph{multi-task learning} (MTL) to simultaneously predict both binding and separation energies, 
achieving a world-leading precision of about 140\,keV; {\em n.b.}\ this model uses the WS4 predictions as important inputs.  
ML models have yet to substantially improve upon analytic-model predictions for charge radii. 
We note that the precision needed for understanding fundamental phenomena in nuclear (astro)physics, are below 100\,keV for binding energies and 0.01\,fm for charge radii;{\em e.g.} for $r$-process nucleosynthesis see ~\cite{Martin:2015xql, Mumpower:2015ova}.

A major drawback of the previous ML-based approaches is that they lack interpretability. 
The nuclear shell model is intuitive and surprisingly accurate given its simplicity. 
We propose an ML-based approach designed to learn a task-independent representation of the nucleus, analogous to the human-learned nuclear shell model picture. 
To decouple the nuclear representation from task-specific model features, we train to predict a variety of nuclear properties in an MTL setting, all based on the same nuclear representation model, referred to as Nuclear Co-Learned Representations (\ainuc). 
By exploiting all information available about nuclei,  we show that we are able to build a more meaningful model and improve prediction quality that achieves state-of-the-art performance. 
More importantly, as shown in  Fig.~\ref{fig:nuc-emb} we can clearly identify the most important aspects of the nuclear shell model in the latent representations of the model that are shared between all tasks, such as the shell structure, including the well-known magic numbers, and the Pauli Exclusion Principle.

\section{More Tasks, More Information}

Improving generalization of a prediction task can be achieved in many ways, 
including by obtaining more or more precise data. %
However, collecting such data is often expensive, and in the case of nuclear data technologically challenging or even infeasible. 
Another avenue is exploiting the joint information between different known and measured properties of the nucleus to increase the effective amount of data available for any specific prediction task. 
This can be achieved by training jointly on all tasks.
To exploit data correlations over multiple tasks, a prediction model needs to process data through dependent channels.
For neural networks, that most often means sharing weights and layers between tasks.
To what degree a model should share processing channels between tasks depends on how similar those tasks are.
In the limit of tasks with no joint information, there is no evident benefit to joint training.

In the MTL literature there exist many ways to either manually or automatically control information sharing \cite{MTLbible}.
In order to introduce and motivate choices made for this MTL task, we first show proof of concept with a toy model.
The task is to predict the result of the following arithmetic binary functions of 
two input real numbers: %
$a+b$, $|a-b|$, $(a + b)^{3/2}$, $\log (a+b+1)$, and exp$(-\sqrt{a+b}/5)$. 
The input numbers are treated like symbols or tokens in a language task by embedding them randomly into a high-dimensional space. 
Once embedded, they lose all their relational properties. 
The representation vectors are trainable parameters during the learning process that allows the architecture to arrange them in a beneficial way.
Schematically, the architecture for this toy-model regression processes inputs is as follows:
embed both numbers,
concatenate these embeddings, 
then send the resulting vector through a stack of residual blocks. 

\begin{figure}[t!]
\centering
\includegraphics[width=0.45\textwidth, trim={0 0 0 .8cm},clip]{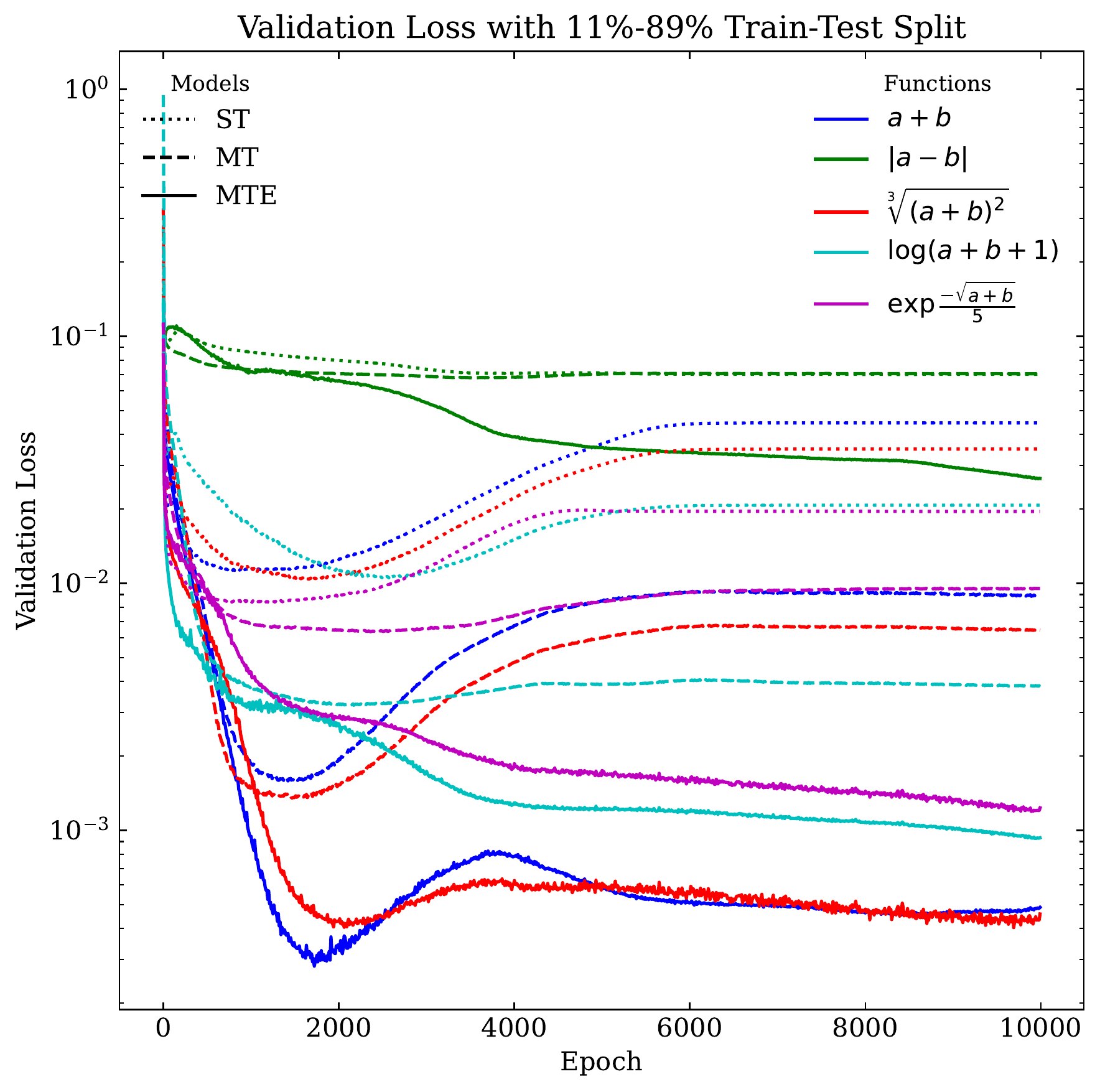}
\caption{For the toy-model arithmetic problem, the multi-task with embeddings (MTE) model generalizes better than the multi task (MT) and single task (ST) models on all tasks in the regime of limited data. 
Here, only 400 examples for each task are used, 10\% for training and the rest for testing.  }
\label{fig:toy_model}
\vspace{-1em}
\end{figure}

In the regime of limited data, we find that using multiple tasks improves generalization on all tasks simultaneously.
Figure~\ref{fig:toy_model} shows the performance obtained when training on individual targets (single-task, ST) versus joint training on all tasks at once (multi-task, MT), where the latter is superior for all tasks.
Here, the MT model has the same layout as the ST models except that the last layer of the model has an output for each task.
Note how this architecture hard-codes a high degree of shared computation between tasks. 
Regardless of the regression target, input numbers undergo the same transformations up to the last layer.

\begin{figure}[t!]
\centering
\includegraphics[width=0.49\textwidth]{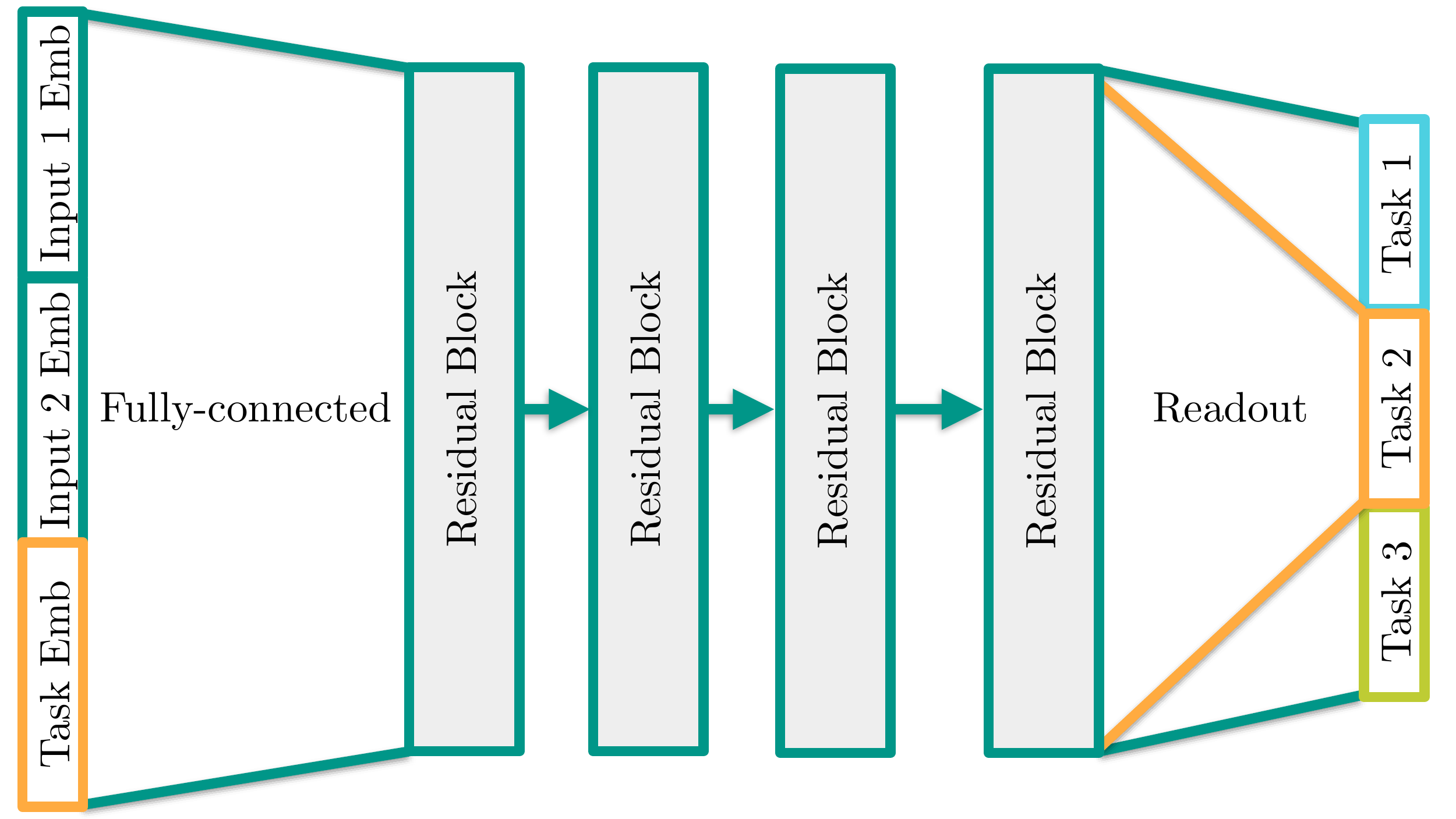}
\caption{MTE architecture: input-data and task embeddings are concatenated, projected, and passed through a sequence of residual blocks. Different readout heads output predictions for each task. The fact that the same input-data embeddings are used for all tasks encourages structure formation in the embedding space which encodes task-independent information.}
\label{fig:model}
\end{figure}

Next, we show that far superior performance is achieved by using a different approach to multi-tasking. 
Traditionally in MTL, architectures define hyperparameters to control the degree of weight sharing for different tasks. 
Taking inspiration from language models---which do implicit multi-tasking by conditioning on the task as an additional input---we instead treat the binary operations as trainable embeddings in the same way as the input numbers, and concatenate all three embedding vectors for processing through the model. 
As in the MT model, the last layer has one output per task.
Our approach lets the model decide to which degree tasks are similar and how computations should be shared 
by moving the task embeddings closer together or farther apart in latent space. 
Figure \ref{fig:toy_model} shows that this model---labelled MTE, shown schematically in Fig.~\ref{fig:model}---yields superior performance over both single-target and naive multi-target training.

\begin{figure*}[t!]
    \centering
\includegraphics[width=0.45\textwidth]{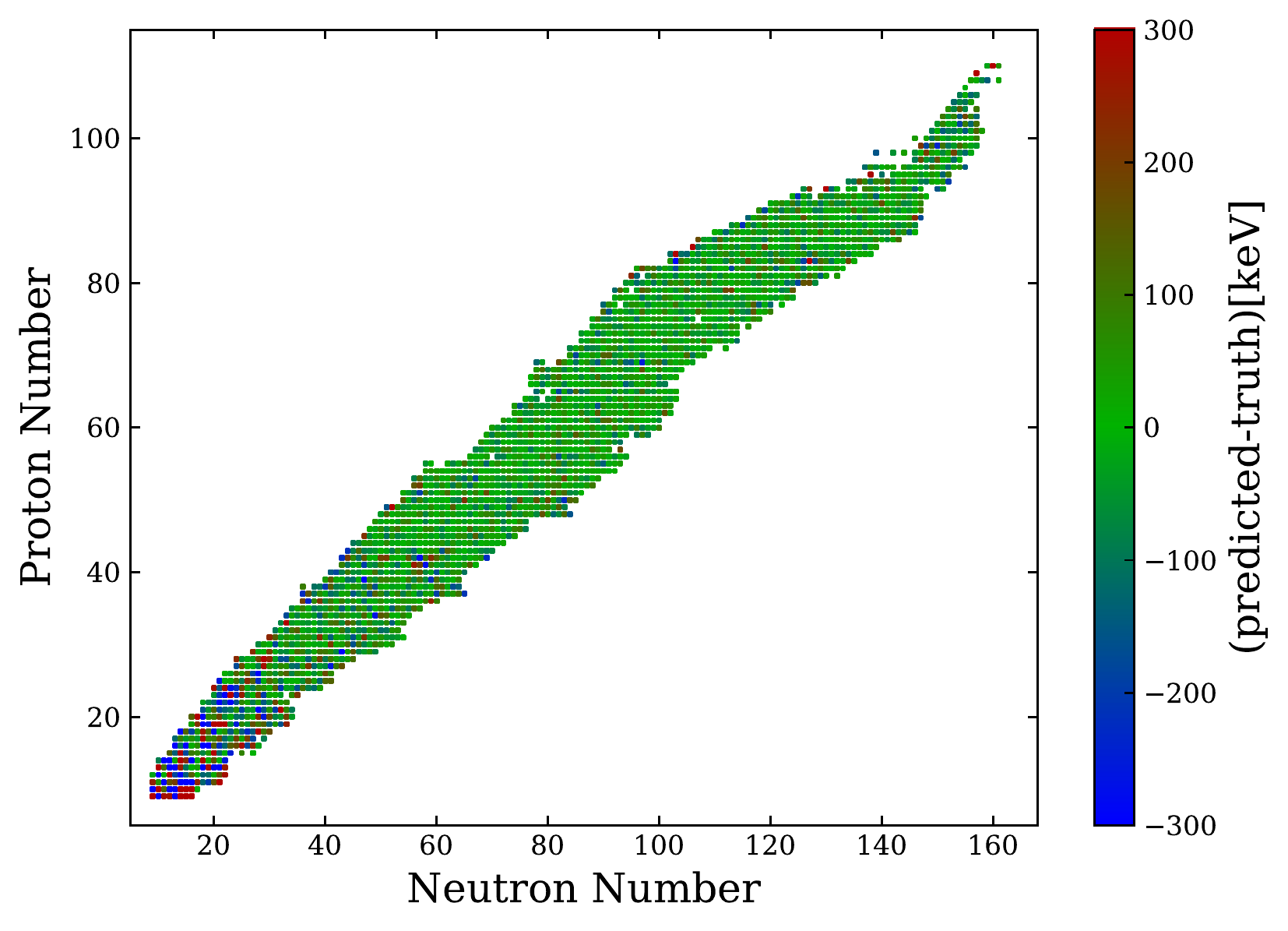} 
\includegraphics[width=0.45\textwidth]{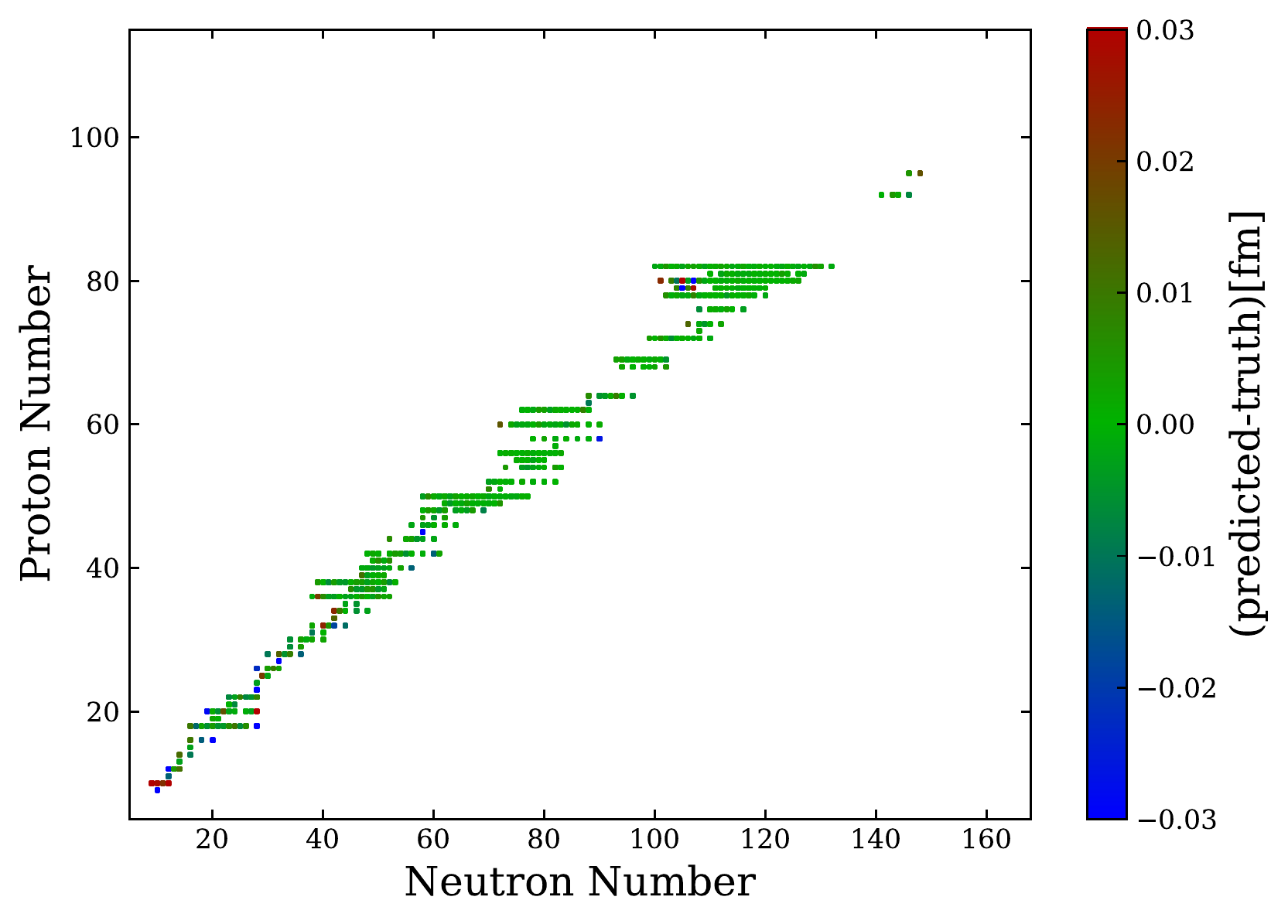} \vspace{-.4em}
    \caption{Residuals for \ainuc predictions compared to experimental measurements for (left) binding energy (keV) and  (right) charge radius (fm). 
    The vast majority of our predictions are highly precise, especially from the $5^{\rm th}$ shell up ($N,Z > 28$) where the RMS of our binding energy predictions is only 79\,keV {\em cf.}\ the binding energies themselves which are  $\mathcal{O}(7000)\cdot(Z+N)$\,keV. }
    \label{fig:nuc-validation}
\end{figure*}

\section{Predicting Nuclear Properties}

We now use the MTE approach presented in the previous section to predict nuclear properties.
We base our training on the AME2020 dataset~\cite{Wang:2021xhn} for the energy-based quantities and on the compilation of \cite{Angeli:2013epw} for the charge radii, in both cases using all well-measured nuclei with $Z,N > 8$.\footnote{More details on the dataset, model, and validation procedure are provided in an appendix.} 
The only inputs are three embeddings: proton and neutron number, $Z$ and $N$, and the task.
We choose the following nuclear properties to predict:
binding energy per nucleon; 
charge radius; 
neutron and proton separation energies, 
defined as the energy required to remove one neutron or proton from the nucleus; 
and the energy available for $\alpha$, $\beta$, electron-capture (EC), and $\beta$-$n$ decays. 
All targets are scaled to be contained in the interval [0,1].
In addition, for binding energies we predict the residuals relative to the semi-empirical Bethe-Weizs{\"a}cker (BW) mass formula, a direct consequence of the famous liquid-drop model of the nucleus~\cite{Weizsacker:1935bkz,Bethe:1936zz}.
The BW formula itself has a prediction precision of about $3~\rm MeV$, which is 30 times worse than our target performance.
Removing the BW predictions does not greatly affect the prediction performance; however, it does affect the interpretability (discussed later).

When evaluating our models, we must avoid prediction biases such as correlations between the separation energies of a nucleus and the binding energies of its neighboring nuclei, which could be used to artificially obtain excellent performance~\cite{Wu:2022nnc}; {\em e.g.}, it is possible to directly obtain (and memorize) the binding energy, $B(Z,N)$, from $B(Z,N) = B(Z,N-1)+S_n(Z,N)$ where $S_n$ denotes the neutron separation energy. 
Therefore, we train many independent models (100-fold cross-validation), where for each, we withhold 1\% of the nuclei from training to be used for validation. 
In addition, we also remove the separation energies of the neighboring nuclei for each nucleus in a validation sample.

All inputs are embedded into a 1024-dimensional space.
The model consists of one linear layer that takes the concatenated embeddings in $3\times1024$ dimensions and maps them back to 1024 dimensional space, 4 residual blocks with 2 layers of width 1024 each,  and one final output layer with one output for each task. We use SiLU activation after each linear layer.
This results in a total of $11\,869\,194$ parameters.
We train each model for 50\,000 epochs with learning rate of 0.01, weight decay of 0.01, and a batch size of 4096. 
The learning rate is cosine-annealed to $10^{-5}$ during the training.

Figure~\ref{fig:nuc-validation} shows the residuals for our predictions for binding energy and charge radius.
The vast majority of our predictions are highly precise, especially for larger nuclei. 
The root-mean-square (RMS) of the prediction performance of our model on the various tasks is as follows:
130\,keV for binding energy; 
0.011\,fm for charge radius; 
130\,keV and 140\,keV for neutron and proton separation energies; 
and 139\,keV, 180\,keV, 179\,keV, 196\,keV for the energy available for $\alpha$, $\beta$, EC, and $\beta$-$n$ decays.
We believe that these results establish a new state of the art. 
In addition, we note that to get competitive performance, previous works considered more human-designed inputs, for instance binding energy predictions of high-precision analytical models~\cite{Niu:2022gwo,Wu:2022nnc}. 
We could likely improve our results by also including such inputs, especially for smaller nuclei; however, our focus here was to demonstrate the ability to self learn the nuclear shell structure in an interpretable way.

\section{Gaining Understanding via Embeddings}

The deep learning literature suggests that the crucial phenomenon of \emph{capability emergence} arises as one scales up the number of parameters, dataset size, and compute across modalities, and in particular, this phenomenon is extremely important in language models \cite{gpt3, emergent_caps}. 
A plausible explanation for the performance gains attributed to this phenomenon is the emergence of specialized neural circuits and higher-quality representations. 
Similarly, since we use the same nuclear embeddings to tackle a range of tasks, we expect our training procedure to lead to the emergence of physically-meaningful representations. 
Furthermore, if most of the information stored in the embeddings lies on a low-dimensional manifold, then the learned solution could be interpretable, allowing us to gain both confidence in and understanding from the model.

To study the embeddings, we project them down into a low-dimensional space via a principal component analysis (PCA). 
Figure \ref{fig:nuc-emb}~(left) shows the first two principle components of the neutron embeddings at an early stage in the training. 
We clearly see that the model has learned to separate the even and odd values of $N$ (vertically on the plot).
Recall that $N$ is not given to the machine as an integer, that fact is obscured by providing $N$ as a randomly initialized vector in a 1024-dimensional space. 
This clear separation of even and odd is a striking statement that it has learned the consequences of the Pauli Exclusion Principle. 
Furthermore, already at this early stage of training we see {\em chains} of connected numbers arise with clear breaks between the chains.
The break points in the even chains are the well-known magic numbers for $N$, {\em i.e.}\ these break points occur where the nuclear shells become full.
Therefore, we conclude that the earliest stages of learning correspond to building embeddings that represent the famous nuclear shell model! 

Figure \ref{fig:nuc-emb}~(right) shows the $N$ embeddings at the end of training, where
the clear even-odd split due to the Pauli principle remains. %
Interestingly, we see that the chains of numbers attributed to the shell structure have evolved into 3-dimensional spirals, where the numbers from each shell are roughly equidistant around a circle in the 2-dimensional subspace with the different shells separated along the third PCA dimension. 
There are many potential explanations for this sturcture, though more detailed study is required to understand this structure.
Nevertheless, this spiral structure appears crucial to \ainuc being able to provide state-of-the-art precision for nuclear predictions. 
Finally, we note that the final embeddings and the observable predictions obtained both with and without the use of the imprecise BW binding-energy predictions as inputs are similar.  
The primary benefit of using the BW predictions is the clear emergence of the nuclear shell model early in the training. %

\section{Discussion and Future Directions}

\ainuc achieves state-of-the-art performance predicting nuclear observables using a multi-task approach with shared representations. 
Amazingly, these learned representations exhibit the prominent emergence of 
the most crucial aspects of the nuclear shell model, 
suggesting the model is capable of capturing the underlying physical principles.
Future directions include more detailed study of the embedding structures, and more generally how to encourage embedding as much information as possible into an interpretable low-dimensional manifold. 
On the nuclear-physics side, \ainuc could potentially be used to make accurate predictions about many exciting topics in nuclear (astro)physics, including $r$-process nucleosynthesis~\cite{Burbidge:1957vc}, the nuclear neutron skin and its consequences for the structure of neutron stars~\cite{Brown:2000pd, Horowitz:2000xj, Gandolfi:2011xu}, the exploration of the boundaries of the nuclear landscape~\cite{Erler2012TheLO}, and of exotic phenomena such as halo nucleii~\cite{Nortershauser:2008vp} and shape coexistence~\cite{Heyde:1983zz,Wood:1992zza}, and the CKM unitarity puzzle~\cite{Seng:2018yzq,Belfatto:2023tbv,Seng:2022inj}.

\section*{Broader Impact} 

A generalized view of the method proposed here is that the scientific endeavor itself is a representation-learning problem. 
Nature presents us with high-dimensional seemingly unorganized data, from which scientists attempt to find low-dimensional representations of the relevant information. 
Once a low-dimensional representation is found, a model can be built that is not only precise but also trustworthy because we can understand how its predictions arise.
In many scientific applications, ML models can easily achieve superior performance but they lack interpretability.  
Our method re-formulates the ML approach to science in a way that is more aligned with how traditional human-led science works.
We showed that if most of the information stored in the embeddings lies on a low-dimensional manifold, then the ML solution could be interpretable, allowing us to gain both confidence in and understanding from the ML model. 
In this way, we not only obtain more precise models---but we can potentially learn from the machine.
We believe that our approach could be applied to any scientific problem where understanding what the ML model has learned is desired. 
Finally, we do not foresee any ethical concerns associated with our work.

\section*{Acknowledgement}
We are grateful to Yotam Soreq and Ben Ohayon for the very useful comments.
This work was supported by NSF grant PHY-2019786 (The NSF AI Institute for Artificial Intelligence and Fundamental Interactions, http://iaifi.org/). ST is also supported by the Swiss National Science Foundation - project n. P500PT\_203156.

\section*{Appendix: Additional NuCLR Details}
\label{sec:app}

\textit{Dataset}: 
The dataset is built from the nuclei in the AME2020 collection, which contains both experimentally measured nuclear data and model predictions for as-yet-unobserved nuclei. 
Only experimentally measured nuclei with $Z,N > 8$ are considered when evaluating model performance; however, predictions for unobserved nuclei are used in the training, which helps in regions where there is little experimental information. 
In total, about 3k nuclei are used in the training and testing. 
Note that most nuclei do not have measurements for all of the properties that we predict, {\em e.g.}, only around one third of the AME2020 nuclei posses a charge-radius measurement. 
For more details, the AME2020 sample can be inspected interactively at \url{https://www-nds.iaea.org/relnsd/vcharthtml/VChartHTML.html}.

\textit{Model}:
Figure~\ref{fig:model} shows a schematic diagram of the model that takes as input the embedding vectors for proton and neutron number, along with the specific prediction task.
During a forward pass, all 3 inputs (proton number, neutron number, task ID) are embedded in a $1024$-dimensional space using a learned embedding function.
These are then concatenated to form a $3\times1024$-dimensional vector. 
This vector is passed through a set of residual blocks of size $1024$, ending in one readout linear layer that transforms the $1024$-dimensional penultimate activation into a vector with entries for each task. 
Thus, branching for multi-tasking happens on two levels: (1) the task embedding and (2) the last-layer routing.
One data point is defined as ($Z$, $N$, Task ID), and the loss for that point considers only the task-relevant output of the network.

\textit{Cross-Validation}: 
We use 100-fold cross validation because neighboring nuclei must be excluded in the training (see \cite{Wu:2022nnc}, and discussion in the main text above).
For each nucleus in the validation set, all neighbor energies are removed from the training sample, leading to $4\mbox{--}5\times$ validation set cardinality removed from the training data in each fold. 
This requires that each validation set be kept small to allow for a large enough training sample. 
Thus, a large number of folds are needed so that each nucleus appears in at least one validation sample. 
We note that the previous state-of-the-art model for predicting nuclear properties \cite{Wu:2022nnc} does a much higher number of folds; they train one model per nucleus.

\bibliography{example_paper}
\bibliographystyle{synsml2023}

\end{document}